\documentclass[10pt,aps,prd,nofootinbib,superscriptaddress,floatfix]{revtex4}
\usepackage[utf8]{inputenc}
\usepackage{amsmath}
\usepackage{amssymb}
\usepackage{graphicx}
\usepackage[bookmarks=false, breaklinks=false,pdfborder={0 0 1},backref=section,colorlinks=false]{hyperref}
\hypersetup{ colorlinks,bookmarksopen,bookmarksnumbered,linkcolor=blu,pdfstartview=FitH,urlcolor=rossos,citecolor=rossos}

\usepackage{color}
\usepackage{textcomp}
\usepackage{amsfonts}
\usepackage{graphics}
\usepackage{epstopdf}
\usepackage{slashed}
\usepackage{multirow}
\usepackage{adjustbox}
\usepackage{lipsum}
\usepackage{rotating}
\usepackage{xcolor}
\usepackage{wrapfig}
\usepackage{epsfig}
\usepackage{ulem}
\usepackage{tikzsymbols}
\usepackage{tikz}

\definecolor{rosso}{cmyk}{0,1,1,0.4}
\definecolor{rossos}{cmyk}{0,1,1,0.55}
\definecolor{rossoc}{cmyk}{0,1,1,0.2}
\definecolor{blu}{cmyk}{1,1,0,0.3}
\definecolor{blus}{cmyk}{1,1,0,0.6}
\definecolor{bluc}{cmyk}{1,1,0,0.1}
\definecolor{verde}{cmyk}{0.92,0,0.59,0.25}
\definecolor{verdec}{cmyk}{0.92,0,0.59,0.15}
\definecolor{verdes}{cmyk}{0.92,0,0.59,0.4}
\definecolor{RED}{rgb}{1,0,0}
\definecolor{BLUE}{rgb}{0,0,1}

\definecolor{lime}{HTML}{A6CE39}
\DeclareRobustCommand{\orcidicon}{
	\begin{tikzpicture}
	\draw[lime, fill=lime] (0,0)
	circle [radius=0.2]
	node[white] {{\fontfamily{qag}\selectfont \tiny ID}};
	\draw[white, fill=white] (-0.0625,0.095)
	circle [radius=0.007];
	\end{tikzpicture}
	\hspace{-2mm} }
\foreach \x in {A, ..., Z}{\expandafter\xdef\csname orcid\x\endcsname{\noexpand\href{https://orcid.org/\csname orcidauthor\x\endcsname}
			{\noexpand\orcidicon}} }

\definecolor{lime}{HTML}{A6CE39}
\foreach \x in {A, ..., Z}{\expandafter\xdef\csname orcid\x\endcsname{\noexpand\href{https://orcid.org/\csname orcidauthor\x\endcsname}
			{\noexpand\orcidicon}} }

\newcommand{\arXivold}[1]{\texttt{arXiv:#1}}

\makeatother

\begin{document}
\title{Novel and Updated Bounds on Flavor-Violating Z Interactions in the
Quark Sector}
\author{Fayez Abu-Ajamieh\orcidA{}}
\email{fayezajamieh@iisc.ac.in}

\affiliation{Formerly: Center for High Energy Physics; Indian Institute of Science;
Bangalore; India}
\author{Amine Ahriche\orcidB{}}
\email{ahriche@sharjah.ac.ae}

\affiliation{Department of Applied Physics and Astronomy; University of Sharjah;
P.O. Box 27272 Sharjah; UAE}
\author{Suman Kumbhakar\orcidC{}}
\email{kumbhakar.suman@gmail.com}

\affiliation{Department of Physics, University of Calcutta, 92 Acharya Prafulla
Chandra Road, Kolkata 700009, India}
\author{Nobuchika Okada\orcidA{}}
\email{okadan@ua.edu}

\affiliation{Department of Physics and Astronomy; University of Alabama;\\
 Tuscaloosa; Alabama 35487; USA}
\begin{abstract}
We derive bounds on the flavor-violating (FV) couplings of the $Z$
boson to quarks and present future sensitivity projections. Our analysis
shows that the current bounds on the FV couplings are $\mathcal{O}(10^{-9})$
for the $Z$ couplings to $cu$ and $sd$, $\mathcal{O}(10^{-7})$
for $bd$, $\mathcal{O}(10^{-6})$ for $bs$, and $\mathcal{O}(10^{-3})$
for $tu$ and $tc$. Overall, low-energy flavor experiments provide
significantly stronger constraints on these FV couplings than current
collider searches. 
\end{abstract}
\maketitle

\section{Introduction}

The Standard Model (SM) of particle physics stands as a monumental
achievement, successfully accounting for a vast range of experimental
results with remarkable precision. Its predictions have been continually
affirmed by data from the Large Hadron Collider (LHC), including the
landmark discovery of the Higgs boson~\cite{ATLAS:2012yve,CMS:2012qbp}.
Despite its success, the SM leaves several fundamental questions unanswered.
These include the nature of dark matter and dark energy, the origin
of the matter--antimatter asymmetry in the universe, the strong CP
problem, and the flavor structure of fermions. Such shortcomings suggest
that the SM should be regarded as a low-energy effective field theory
(EFT), valid below some high-energy scale $\Lambda$, beyond which
new physics (NP) is expected to emerge. In this context, understanding
the flavor structure of quarks is deeply tied to non-perturbative
QCD dynamics. Recent advances in quark models, as reviewed in~\cite{Nefediev:2025vmo},
help illuminate fundamental QCD phenomena, such as chiral symmetry
breaking and hadron structure, and provide a complementary motivation
for exploring flavor-violating interactions beyond the SM.

In the effective field theory (EFT) paradigm, new physics is incorporated
in a model-independent way through higher-dimensional operators constructed
from SM fields. Widely used EFT frameworks include the Standard Model
EFT (SMEFT)~\cite{Buchmuller:1985jz,Grzadkowski:2010es}, the Higgs
EFT (HEFT) (see~\cite{Brivio:2017vri} and the references therein),
and formulations in terms of deviations in observed couplings~\cite{Chang:2019vez,Abu-Ajamieh:2020yqi,Abu-Ajamieh:2021egq,Abu-Ajamieh:2022ppp}.
In particular, recent overviews of Higgs physics within the EFT approach,
such as~\cite{Grober:2025eyl}, highlight how deviations in Higgs
couplings can probe new physics scales and complement searches for
flavor-violating interactions.

Among various approaches to probing physics beyond the Standard Model
(BSM), one particularly promising direction is the search for flavor-violating
(FV) processes, especially in neutral current interactions. While
the SM forbids flavor-changing neutral currents (FCNCs) at tree level~\cite{Glashow:1970gm},
it is straightforward to generalize flavor-conserving neutral current
interactions, such as those of the photon, $Z$ boson, or Higgs boson,
to include FV couplings. However, making the QED coupling flavor-dependent
would result in charge dequantization, which is tightly constrained~\cite{Foot:1992ui,Giunti:2014ixa,Raffelt:1999gv,Abu-Ajamieh:2023txh},
leaving only the $Z$ and Higgs bosons as viable candidates for carrying
FV interactions.

FV interactions involving the Higgs boson have been extensively studied
in both the quark and lepton sectors~\cite{Bjorken:1977vt,McWilliams:1980kj,Shanker:1981mj,Barr:1990vd,Babu:1999me,Diaz-Cruz:1999sns,Han:2000jz,Blanke:2008zb,Casagrande:2008hr,Giudice:2008uua,Aguilar-Saavedra:2009ygx,Albrecht:2009xr,Buras:2009ka,Agashe:2009di,Goudelis:2011un,Arhrib:2012mg,McKeen:2012av,Azatov:2009na,Blankenburg:2012ex,Kanemura:2005hr,Davidson:2010xv,Harnik:2012pb,Fernandez:2009vr,Aranda:2009cd,Aranda:2010qc,Abu-Ajamieh:2022nmt,Abu-Ajamieh:2023qvh,Abu-Ajamieh:2025jsz}.
In contrast, FV couplings of the $Z$ boson, especially those involving
quarks, have received comparatively little attention~\cite{Brignole:2004ah,Davidson:2012wn,Goto:2015iha,Kamenik:2023hvi,Jueid:2023fgo,Calibbi:2021pyh,Chivukula:2002ry,Erler:2009jh,Aranda:2010cy,Murakami:2001cs,Altmannshofer:2016brv},
with most existing studies focusing on the lepton sector.

A comprehensive analysis of FV $Z$ interactions with leptons was
presented in~\cite{Abu-Ajamieh:2025vxw}, which established updated
and novel bounds on these couplings. In this work, we extend that
analysis to the quark sector. Building upon the framework developed
in~\cite{Abu-Ajamieh:2025vxw}, which we briefly review in Section~\ref{sec:framework},
we derive current bounds and future projections on FV $Z$ couplings
to quarks. To our knowledge, this constitutes the first dedicated
and comprehensive study of this kind.

We obtain constraints on FV $Z$ interactions involving quarks from
several experimental inputs, including direct searches, meson oscillations,
rare top decays (e.g., $t \to Zq$), electroweak precision observables
(EWPOs), and leptonic decays of neutral mesons containing different
quark flavors. Our analysis shows that bounds from EWPOs are the weakest,
at the level of $\mathcal{O}(0.1)$, while direct search limits are
slightly stronger, around $\mathcal{O}(10^{-2})$. Constraints from
top decays reach the $\mathcal{O}(10^{-3})$ level, and meson oscillations
provide the most stringent bounds, ranging from $\mathcal{O}(10^{-6})$
to $\mathcal{O}(10^{-9})$. Bounds from leptonic decays of neutral
mesons fall between $\mathcal{O}(10^{-2})$ and $\mathcal{O}(10^{-7})$.
We also provide future sensitivity projections from the proposed Future
Circular Collider electron-positron (FCC-ee) and International Linear
Collider (ILC) experiments. Interestingly, the current constraints
from meson data are already stronger than the projected bounds from
the ILC for some couplings, while the FCC-ee is expected to improve
sensitivity by up to two to three orders of magnitude.

The structure of this paper is as follows: In Section~\ref{sec:framework},
we briefly review the theoretical framework used to parameterize FV
$Z$ couplings. Section~\ref{sec:bounds} presents the derivation
and discussion of the bounds. We summarize our conclusions in Section~\ref{sec:conclusion}.

\section{The Framework~\label{sec:framework}}

In this section, we briefly go over the framework developed in~\cite{Abu-Ajamieh:2025vxw}.
In the SM, the $Z$ interaction to fermions is flavor-conserving and
is given by

\begin{equation}
\mathcal{L}_{Zf\bar{f}}^{\text{SM}}=-Z_{\mu}\sum_{i}\bar{f}_{i}\gamma^{\mu}(g_{L}^{f}P_{L}+g_{R}^{f}P_{R})f_{i},\label{eq:Sm_Z_cpoupling}
\end{equation}
where $g_{L}^{f}=\frac{g}{\cos\theta_{W}}(T_{fL}^{3}-Q_{f}\sin^{2}\theta_{W})$
and $g_{R}^{f}=\frac{-g}{\cos\theta_{W}}(Q_{f}\sin^{2}\theta_{W})$,
with both quantities being real in the SM. By utilizing a bottom-up
approach, it is straightforward to incorporate FV in Eq. (\ref{eq:Sm_Z_cpoupling})
by promoting $g_{L,R}^{f}$ to become non-diagonal, possibly complex,
matrices,\footnote{Note that eq.(\ref{eq:FV_Z_cpoupling}) is self-Hermitian as we consider
$g_{L,R}^{ij}=g_{L,R}^{ji*}$. To see this, we have $\sum_{i,j}(\bar{q}_{i}\gamma_{\mu}(g_{L}^{ij}P_{L}+g_{R}^{ij}P_{R})q_{j})^{*}=\sum_{i,j}\bar{q}_{j}(g_{L}^{ij*}P_{R}+g_{R}^{ij*}P_{L})\gamma_{\mu}q_{i}=\sum_{i,j}\bar{q}_{j}\gamma_{\mu}(g_{L}^{ij*}P_{R}+g_{R}^{ij*}P_{L})q_{i}=\sum_{i,j}\bar{q}_{i}\gamma_{\mu}(g_{L}^{ji*}P_{R}+g_{R}^{ji*}P_{L})q_{j}=\sum_{i,j}\bar{q}_{i}\gamma_{\mu}(g_{L}^{ij}P_{L}+g_{R}^{ij}P_{R})q_{j}$.} 
\begin{equation}
\mathcal{L}_{Zf\bar{f}}^{\text{FV}}=-Z_{\mu}\sum_{i,j}\overline{f}_{i}\gamma^{\mu}(g_{L}^{ij}P_{L}+g_{R}^{ij}P_{R})f_{j},\label{eq:FV_Z_cpoupling}
\end{equation}
with the diagonal elements of the matrices $g_{L,R}^{ij}$ corresponding
to the SM flavor-conserving couplings 
\begin{align}
g_{L}^{\text{SM}}=g_{L}^{ii}\equiv g_{L}^{f},\hspace{5mm}g_{R}^{\text{SM}}=g_{R}^{ii}\equiv g_{R}^{f}.\label{eq:SM_gLR}
\end{align}

In general, the $g_{L,R}^{ij}$ matrices are complex Hermitian, however,
for simplicity, we assume them real symmetric as we did in~\cite{Abu-Ajamieh:2025vxw}.
The off-diagonal terms could arise from higher dimensional operators.
For instance, in the dim-6 SMEFT, the following operators contribute
(in the notation of the Warsaw basis~\cite{Grzadkowski:2010es})
\begin{equation}
\mathcal{Q}_{\text{SMEFT}}^{\text{dim}-6}=\{Q_{\varphi q}^{(1)},Q_{\varphi q}^{(3)},Q_{\varphi u},Q_{\varphi d},Q_{\varphi ud},Q_{uG},Q_{uW},Q_{uB},Q_{dG},Q_{dW},Q_{dB}\}.
\end{equation}

It is possible to construct $g_{L,R}^{ij}$ with SM-diagonal entries
and BSM off-diagonal entries without significant fine-tuning. This
was discussed in detail in~\cite{Abu-Ajamieh:2025vxw} for the lepton
sector. The quark sector is not much different than the lepton one,
that is not discussed here. The interested reader is instructed to
refer to~\cite{Abu-Ajamieh:2025vxw} for detail. Explicitly, the
FV Lagrangian reads 
\begin{equation}
\mathcal{L}_{Zq\bar{q}}^{\text{BSM}}=-Z^{\mu}\sum_{i,j}\bar{u}_{i}\gamma_{\mu}(g_{u_{L}}^{ij}P_{L}+g_{u_{R}}^{ij}P_{R})u_{j}-Z^{\mu}\sum_{i,j}\bar{d}_{i}\gamma_{\mu}(g_{d_{L}}^{ij}P_{L}+g_{d_{R}}^{ij}P_{R})d_{j},
\end{equation}
where $u_{i}=\{u,c,t\}$ and $d_{i}=\{d,s,b\}$. This is our master
formula that we will use to set bounds on the FV matrices $g_{L,R}^{i,j}$.

\section{Bounds on The FV $Z$ Interactions~\label{sec:bounds}}

In this section, we discuss the experimental bounds on the FV $Z$
couplings to quarks in detail. In all our calculation, we shift all
bounds to be at $90\%$ CL to be consistent. The bounds and future
projections are summarized in Table (\ref{table1}) and Figures \ref{fig3}.

\begin{table}[t]
\centering \vspace{1mm}
 \tabcolsep7pt%
\begin{tabular}{l|c|c|c}
\hline 
\textbf{Channel} & \textbf{Couplings} & \textbf{Bounds} & \textbf{Projections} \tabularnewline
\hline 
\hline 
Direct Searches & $\sqrt{|g_{L}^{cu}|^{2}+|g_{R}^{cu}|^{2}}$ & $<4.96\times10^{-2}$ & - \tabularnewline
Direct Searches & $\sqrt{|g_{L}^{bs}|^{2}+|g_{R}^{bs}|^{2}}$ & $<4.96\times10^{-2}$ & $10^{-5}$ \tabularnewline
Direct Searches & $\sqrt{|g_{L}^{bd}|^{2}+|g_{R}^{bd}|^{2}}$ & $<4.96\times10^{-2}$ & $10^{-4}$ \tabularnewline
Direct Searches & $\sqrt{|g_{L}^{sd}|^{2}+|g_{R}^{sd}|^{2}}$ & $<4.96\times10^{-2}$ & - \tabularnewline
$D^{0}-\bar{D}^{0}$ Oscillation & $\sqrt{|g_{L}^{cu}|^{2}+|g_{R}^{cu}|^{2}}$ & $<3.48\times10^{-9}$ & - \tabularnewline
$B^{0}-\bar{B}^{0}$ Oscillation & $\sqrt{|g_{L}^{bd}|^{2}+|g_{R}^{bd}|^{2}}$ & $<1.11\times10^{-7}$ & - \tabularnewline
$B^{s}-\bar{B}^{s}$ Oscillation & $\sqrt{|g_{L}^{bs}|^{2}+|g_{R}^{bs}|^{2}}$ & $<5.32\times10^{-6}$ & - \tabularnewline
$K^{0}-\bar{K}^{0}$ Oscillation & $\sqrt{|g_{L}^{sd}|^{2}+|g_{R}^{sd}|^{2}}$ & $[-4.64,4.64]\times10^{-9}$ & - \tabularnewline
$t \to Zu$ Decay & $\sqrt{|g_{L}^{tu}|^{2}+|g_{R}^{tu}|^{2}}$ & $<4.41\times10^{-3}$ & $10^{-2}-10^{-3}$\tabularnewline
$t \to Zc$ Decay & $\sqrt{|g_{L}^{tc}|^{2}+|g_{R}^{tc}|^{2}}$ & $<6.39\times10^{-3}$ & $10^{-2}-10^{-3}$ \tabularnewline
EWPO & $\sqrt{|g_{L}^{ij}|^{2}+|g_{R}^{ij}|^{2}}$ & $<0.2$ & - \tabularnewline
$D^{0} \to \mu^{+}\mu^{-}$ & $\sqrt{|g_{L}^{cu}|^{2}+|g_{R}^{cu}|^{2}}$ & $<7.44\times10^{-5}$ & -\tabularnewline
$D^{0} \to e^{+}e^{-}$ & $\sqrt{|g_{L}^{cu}|^{2}+|g_{R}^{cu}|^{2}}$ & $<7.76\times10^{-2}$ & -\tabularnewline
$B^{0} \to \tau^{+}\tau^{-}$ & $\sqrt{|g_{L}^{bd}|^{2}+|g_{R}^{bd}|^{2}}$ & $<1.50\times10^{-3}$ & -\tabularnewline
$B^{0} \to \mu^{+}\mu^{-}$ & $\sqrt{|g_{L}^{bd}|^{2}+|g_{R}^{bd}|^{2}}$ & $<5.79\times10^{-6}$ & -\tabularnewline
$B^{0} \to e^{+}e^{-}$ & $\sqrt{|g_{L}^{bd}+|g_{R}^{bd}|^{2}}$ & $<5.40\times10^{-3}$ & -\tabularnewline
$B_{s} \to \tau^{+}\tau^{-}$ & $\sqrt{|g_{L}^{bs}|^{2}+|g_{R}^{bs}|^{2}}$ & $<2.20\times10^{-3}$ & -\tabularnewline
$B_{s} \to \mu^{+}\mu^{-}$ & $\sqrt{|g_{L}^{bs}|^{2}+|g_{R}^{bs}|^{2}}$ & $<1.03\times10^{-5}$ & -\tabularnewline
$B_{s} \to e^{+}e^{-}$ & $\sqrt{|g_{L}^{bs}+|g_{R}^{bs}|^{2}}$ & $<8.6\times10^{-3}$ & -\tabularnewline
$K_{s}^{0} \to \mu^{+}\mu^{-}$ & $\sqrt{|g_{L}^{sd}|^{2}+|g_{R}^{sd}|^{2}}$ & $<4.04\times10^{-6}$ & -\tabularnewline
$K_{s}^{0} \to e^{+}e^{-}$ & $\sqrt{|g_{L}^{sd}+|g_{R}^{sd}|^{2}}$ & $<5.22\times10^{-3}$ & -\tabularnewline
$K_{L}^{0} \to \mu^{+}\mu^{-}$ & $\sqrt{|g_{L}^{sd}|^{2}+|g_{R}^{sd}|^{2}}$ & $<4.96\times10^{-7}$ & -\tabularnewline
$K_{L}^{0} \to e^{+}e^{-}$ & $\sqrt{|g_{L}^{sd}+|g_{R}^{sd}|^{2}}$ & $<7.23\times10^{-6}$ & -\tabularnewline
\hline 
\end{tabular}\caption{{\small{$90\%$ CL bounds and projections on the FV $Z$ couplings
to Quarks. For the bound from the EWPO, we have $\{ij\}=\{tc,tu,cu,bs,bd,sd\}$.}}}
\label{table1} 
\end{table}

\subsection{Direct Searches}

\begin{figure}[t]
\includegraphics[width=0.35\textwidth]{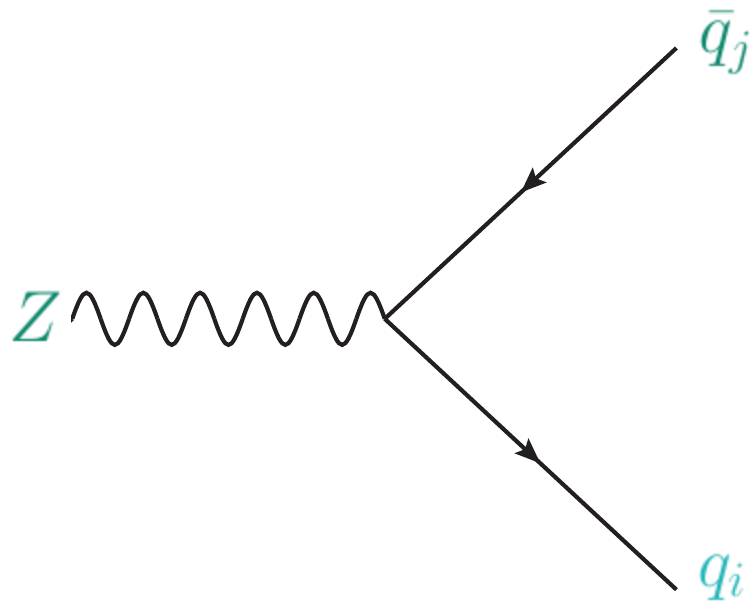} \caption{{\small The FV hadronic decays of the $Z$ gauge boson with $i\protect\neq j$,
that are originated from the interaction in Eq.~(\ref{eq:FV_Z_cpoupling}).}}
\label{figZdecay} 
\end{figure}

The simplest bounds can be extracted from the direct searches of the
hadronic $Z \to q_{i}\bar{q}_{j}$ decay with $i\neq j$ as shown in
Fig.~\ref{figZdecay}. At tree level, the decay width is given by
\begin{equation}
\Gamma(Z \to q_{i}\bar{q}_{j})=\frac{M_{Z}}{12\pi}(|g_{L}^{ij}|^{2}+|g_{R}^{ij}|^{2}),\label{eq:Z_decay}
\end{equation}
where we have neglected the masses of the final state quarks. Bounds
on non-standard hadronic $Z$ decays are obtained from the disagreement
between the SM predictions and the $Z$ hadronic decay width. These
bounds are given by $\text{Br}(Z \to q_{i}\bar{q}_{j})<2.9\times10^{-3}$~\cite{OPAL:2000ufp}
(see also~\cite{Kamenik:2023hvi}), which translates to $\sqrt{|g_{L}^{ij}|^{2}+|g_{R}^{ij}|^{2}}<4.96\times10^{-2}$,
where $\{i,j\}=\{c\bar{u},b\bar{s},b\bar{d},s\bar{d},\bar{c}u,\bar{b}s,\bar{b}d,\bar{s}d\}$.
Notice here that the top quark is not included as $Z$ does not decay
to an on-shell $t$. The FCC-ee is projected to improve the bounds
on the couplings $\{i,j\}=\{b\bar{s},b\bar{d},\bar{b}s,\bar{b}d\}$.
According to~\cite{Kamenik:2023hvi}, the bounds could potentially
improve to $\sqrt{|g_{L}^{bs}|^{2}+|g_{R}^{bs}|^{2}}<10^{-5}$ and
$\sqrt{|g_{L}^{bd}|^{2}+|g_{R}^{bd}|^{2}}<10^{-4}$. The improvement
in the sensitivity is achieved through classifying events according
to how many flavor tagged jets they contain. More specifically, the
inclusion of information about events with light jets can lead to
significant improvement in the sensitivity to $V_{ts}$ and $V_{td}$,
which leads to improving the sensitivity of the bounds. For instance,
in the decay $Z \to bs$, $b$- and $s$-tagging is used, and the events
are distributed into $(n_{b},n_{s})$ bins, with the expected number
of events per bin given by 
\begin{equation}
\overline{N}_{(n_{b},n_{s})}=\sum_{f}p(n_{b},n_{s}|f,\nu)\overline{N}_{f}(\nu),
\end{equation}
with the sum covering the relevant decay channels $f\in\{u\bar{u}+d\bar{d},s\bar{s},c\bar{c},b\bar{b},bs\}$.

\subsection{Meson Oscillation}

\begin{figure}[t]
\includegraphics[width=0.9\textwidth]{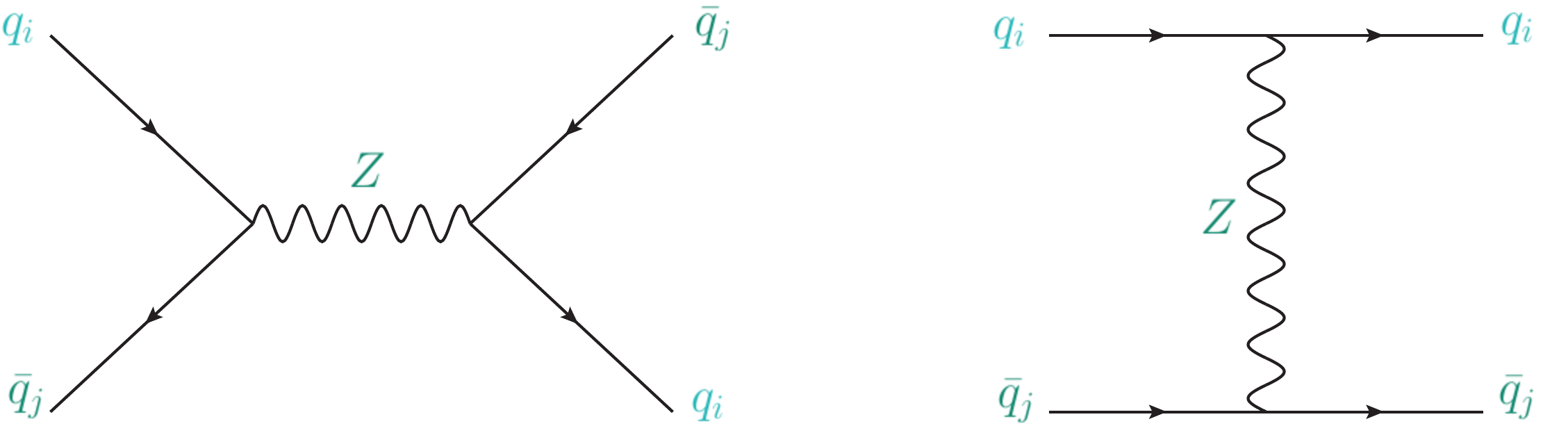} \caption{{\small The FV meson oscillations with $i\protect\neq j$, that are
originated from the interactions in Eq.~(\ref{eq:FV_Z_cpoupling}).}}
\label{fig1} 
\end{figure}

FV $Z$ couplings could contribute to the oscillation of neutral mesons
$B_{d,s}-\bar{B}_{d,s}$, $K^{0}-\bar{K}^{0}$ and $D^{0}-\bar{D}^{0}$.
At tree level, these oscillations proceed via the diagrams shown in
Figure \ref{fig1}. In the non-relativistic limit where the momentum
transfer $q_{Z}^{2}\ll M_{Z}^{2}$, the oscillation cross section
can be approximately expressed as follows

\begin{equation}
\sigma(q_{i}\bar{q}_{j} \to q_{j}\bar{q}_{i})\simeq\frac{M^{2}}{12\pi M_{Z}^{4}}\Big(|g_{L}^{ij}|^{2}+|g_{R}^{ij}|^{2}\Big),\label{eq:Osc_XSection_1}
\end{equation}
where we have dropped terms $\sim\mathcal{O}(m_{i,j}/M_{Z})$ and
have set $\sqrt{s}=M$, the mass of the decaying meson. On the other
hand, the effective Hamiltonian of meson oscillations can be expressed
as~\cite{UTfit:2007eik} 
\begin{equation}
\mathcal{H}_{\text{eff}}=c_{1}(\bar{q}_{jL}^{\alpha}\gamma_{\mu}\bar{q}_{iL}^{\alpha})(\bar{q}_{jL}^{\beta}\gamma_{\mu}\bar{q}_{iL}^{\beta})+\tilde{c}_{1}(\bar{q}_{jR}^{\alpha}\gamma_{\mu}\bar{q}_{iR}^{\alpha})(\bar{q}_{jR}^{\beta}\gamma_{\mu}\bar{q}_{iR}^{\beta}),\label{eq:H_eff}
\end{equation}
which yields the cross section 
\begin{equation}
\sigma(q_{i}\bar{q}_{j} \to q_{j}\bar{q}_{i})\simeq\frac{M^{2}}{48\pi}\Big(|c_{1}|^{2}+|\tilde{c}_{1}|^{2}\Big),\label{eq:Osc_XSection_2}
\end{equation}
where in our case we have $|c_{1}|=|\tilde{c}_{1}|$. Thus, it is
a simple exercise to match eq. (\ref{eq:Osc_XSection_1}) to eq. (\ref{eq:Osc_XSection_2})
and find that $|c_{1}|^{2}=\frac{2}{M_{Z}^{4}}(|g_{L}^{ij}|^{2}+|g_{R}^{ij}|^{2})$.
This allows us to translate the bounds on $c_{1}$ from Table 4 of~\cite{UTfit:2007eik}
into bounds on the entries of the matrices $g_{L,R}^{ij}$.

The first bound comes from the $D^{0}-\bar{D}^{0}$ oscillation, were
the bound is given by $|c_{D}^{1}|<7.2\times10^{-13}\hspace{1mm}\text{GeV}^{-2}$,
which translates to $\sqrt{|g_{L}^{cu}|^{2}+|g_{R}^{cu}|^{2}}<3.48\times10^{-9}$.
The second bound arises from the $B^{0}-\bar{B}^{0}$ oscillation
and is given by $|c_{B_{d}}^{1}|<2.3\times10^{-11}\hspace{1mm}\text{GeV}^{-2}$,
which yields the bound $\sqrt{|g_{L}^{bd}|^{2}+|g_{R}^{bd}|^{2}}<1.11\times10^{-7}$.
The third bound comes from the $B_{s}-\bar{B}_{s}$ oscillation and
is given by $|c_{B_{s}}^{1}|<1.1\times10^{-9}\hspace{1mm}\text{GeV}^{-2}$,
which yields the bound $\sqrt{|g_{L}^{bs}|^{2}+|g_{R}^{bs}|^{2}}<5.32\times10^{-6}$.
Finally, the bounds from the $K^{0}-\bar{K}^{0}$ oscillation are
given by $\text{Re}(c_{K}^{1})\in[-9.6,9.6]\times10^{-13}\hspace{1mm}\text{GeV}^{-2}$
and $\text{Im}(c_{K}^{1})\in[-4.4,2.8]\times10^{-15}\hspace{1mm}\text{GeV}^{-2}$,
which implies that $|c_{K}^{1}|^{2}=|\text{Re}(c_{K}^{1})|^{2}+|\text{Im}(c_{K}^{1})|^{2}\simeq|\text{Re}(c_{K}^{1})|^{2}$.
This bound translates to $\sqrt{|g_{L}^{sd}|^{2}+|g_{R}^{sd}|^{2}}\in[-4.64,4.64]\times10^{-9}$.

\subsection{The $t \to Zq$ Decay}

It is also possible to obtain bounds on the FV $Z$ couplings from
the decay $t \to Zq$, where $q=\{u,c\}$. This decay proceeds at tree
level and the decay width estimation easily gives 
\begin{equation}
\Gamma(t \to Zq)\simeq\frac{m_{t}^{3}}{32\pi M_{Z}^{2}}\Big(|g_{L}^{tq}|^{2}+|g_{R}^{tq}|^{2}\Big)\Big(1-\frac{M_{Z}^{2}}{m_{t}^{2}}\Big)^{2}\Big(1+\frac{2M_{Z}^{2}}{m_{t}^{2}}\Big),\label{eq:top_decay}
\end{equation}
where the $q$ mass can be safely neglected with respect to the top
and Z masses.

From~\cite{ATLAS:2023qzr}, the bounds are given by $\text{Br}(t \to Zu)<6.2\times10^{-5}$
and $\text{Br}(t \to Zc)<1.3\times10^{-4}$, which, using $\Gamma_{t}=1.9$
GeV~\cite{D0:2012hgn} and $m_{t}=172.52$ GeV~\cite{Degrassi:2012ry},
translate into the bounds $\sqrt{|g_{L}^{tu}|^{2}+|g_{R}^{tu}|^{2}}<4.41\times10^{-3}$
and $\sqrt{|g_{L}^{tc}|^{2}+|g_{R}^{tc}|^{2}}<6.39\times10^{-3}$,
respectively.

The proposed International Linear Collider (ILC) is expected to probe
these couplings via several proposed searches. The projected bounds
depend on the proposed COM energy and the type of searches to be conducted
(see~\cite{TopQuarkWorkingGroup:2013hxj} for detail). Here we summarize
the projected searches and extract the corresponding bounds. For anomalous
top quark decay searches at a COM energy of 250 GeV, the projected
bounds read $\text{Br}(t \to Zq)<5(2)\times10^{-4}$, which translate
to $\sqrt{|g_{L}^{tq}|^{2}+|g_{L}^{tq}|^{2}}<1.38\times10^{-2}(8.74\times10^{-3})$,
whereas for the same searches at a COM energy of 500 GeV have projected
bounds of $\text{Br}(t \to Zq)<1.5(1.1)\times10^{-4}$, which translate
to $\sqrt{|g_{L}^{tq}|^{2}+|g_{L}^{tq}|^{2}}<7.57(6.48)\times10^{-3}$.
Finally, searches involving anomalous top quark decays at a COM energy
of 500 GeV have projected bounds of $\text{Br}(t \to Zq)<1.6(1.7)\times10^{-3}$,
which yield the limits $\sqrt{|g_{L}^{tq}|^{2}+|g_{L}^{tq}|^{2}}<2.47(2.55)\times10^{-2}$.
We can see the best of these projections is actually weaker than the
current bounds, which casts doubt on the effectiveness of the ILC
to probe these FV couplings. In addition, collider searches do not
seem to be the optimal venue for probing these couplings, as they
are orders of magnitude weaker than the bounds that can be obtained
from meson oscillations and neutral meson decays as we shall see later
on.

\subsection{The Electorweal Precision Tests}

FV $Z$ couplings to fermions could impact the oblique parameters.
This was evaluated in~\cite{Abu-Ajamieh:2025vxw} for the lepton
sector and the same results apply for the quark sector as well. In
particular, it was found that the corrections to the $STU$ parameters
is given by 
\begin{align}
T & =-\frac{1}{\alpha M_{Z}^{2}}\Pi_{ZZ}^{ij}(0),\label{eq:T_param}\\
S & =-U=\frac{4\cos^{2}\theta_{W}\sin^{2}\theta_{W}}{\alpha M_{Z}^{2}}\Big(\Pi_{ZZ}^{ij}(M_{Z}^{2})-\Pi_{ZZ}^{ij}(0)\Big).\label{eq:SU_param}
\end{align}
where 
\begin{align}
\Pi_{ZZ}^{ij}(0) & \simeq\frac{m_{i}^{2}}{16\pi^{2}}\Big(|g_{L}^{ij}|^{2}+|g_{R}^{ij}|^{2}\Big)\left[1-2\log\Big(\frac{m_{i}^{2}}{\mu^{2}}\Big)\right],\\
\Pi_{ZZ}^{ij}(m_{Z}^{2}) & \simeq\frac{M_{Z}^{2}}{12\pi^{2}}\Big(|g_{L}^{ij}|^{2}+|g_{R}^{ij}|^{2}\Big)\left[\log\Big(\frac{M_{Z}^{2}}{\mu^{2}}\Big)-\frac{5}{3}+i\pi\right],
\end{align}
$m_{i}$ is the mass of the heavier particle in the loop, and $\mu$
is the renormalization scale set to be equal to $M_{Z}$, since the
$STU$ parameters are measured at the $Z$ pole. The latest bounds
on the $STU$ parameters are given by~\cite{ParticleDataGroup:2024cfk}
\begin{align}
S & =-0.04\pm0.10,\hspace{2mm}T=0.01\pm0.12,\hspace{2mm}U=0.05\pm0.09.\label{eq:STU}
\end{align}

As it turns out, only the bound from the $U$ parameter is significant,
and it translate into the limit $\sqrt{|g_{L}^{ij}|^{2}+|g_{R}^{ij}|^{2}}<0.2$,
where $\{ij\}=\{cu,bs,bd,sd,tc,tu\}$. Notice here that we need to
use $\alpha(M_{Z})=1/128.99$. As we can see, the bounds from the
EWPO are not too strong and don't seem to be promising. Therefore,
the EWPOs are not optimal to search for FV.

\subsection{Bounds from the Neutral Meson Decays into Leptons}

\begin{figure}[t]
\includegraphics[width=0.5\textwidth]{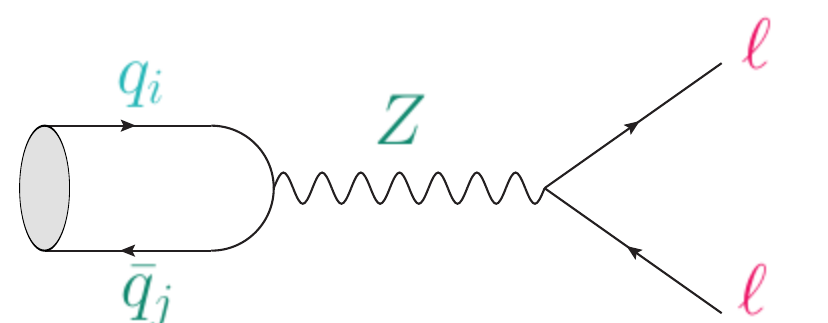} \caption{{\small The FV meson Decays, where the meson is made of different flavor
quark-antiquark; and it decays into same flavor leptons via the $Z$
gauge boson.}}
\label{fig2} 
\end{figure}

One of the novel bounds on the FV $Z$ couplings to leptons obtained
from~\cite{Abu-Ajamieh:2025vxw} is from the meson decays to $\ell_{i}\ell_{j}$,
with $i\neq j$ and with the mesons being made of $\bar{q}q$. The
same type of meson decays can be used to set bounds on the FV $Z$
couplings to quarks, but for mesons made of $\bar{q_{i}}q_{j}$, with
$i\neq j$, decaying to $\ell^{+}\ell^{-}$, as shown in Figure \ref{fig2}.
To this avail, we can use the current algebra formalism. In the SM,
the $Z$ interaction to fermions can be expressed in terms of the
neutral current as follows 
\begin{equation}
\mathcal{L}_{int}=\frac{e}{\sin\theta_{W}}Z_{\mu}J_{Z}^{\mu},\label{eq:Z_current1}
\end{equation}
where 
\begin{equation}
J_{Z}^{\mu}=\sum_{i}\Big(\frac{1}{\cos{\theta_{W}}}\bar{\psi}_{i}\gamma^{\mu}T^{3}\psi_{i}-Q_{i}\frac{\sin^{2}{\theta_{W}}}{\cos{\theta_{W}}}\bar{\psi}_{i}\gamma^{\mu}\psi_{i}\Big),\label{eq:Z_current2}
\end{equation}
and we can easily extend this to include FV by simply introducing
the FV neutral current as follows 
\begin{equation}
J_{Z,\text{FV}}^{\mu}=\sum_{ij}\bar{\psi}_{i}\gamma^{\mu}(g_{L}^{ij}P_{L}+g_{R}^{ij}P_{R})\psi_{j}.\label{eq:FV_Z_current}
\end{equation}

This FV neutral current creates/annihilates neutral mesons made of
different flavors from the QCD vacuum. We are now in a position to
evaluate the decay in Figure \ref{fig2}, however, before we do so
there is an important subtlety that we need to pay attention to. In
the original case that involved $\bar{q}q$ meson decays to $\ell_{i}\ell_{j}$,
$i\neq j$, the decay involved evaluating the matrix element $\left\langle 0\left|J_{Z}^{\mu}\right|M(p)\right\rangle $,
which simply describes the annihilation of the meson (made of quarks
of the same flavor) to become an off-shell $Z$. This matrix element
is generally parameterized through the decay constant of the meson,
which is measured experimentally, however, in the case at hand where
the meson is made of quarks of different species, the matrix element
that needs to be evaluated is $\left\langle 0\left|J_{Z,\text{FV}}^{\mu}\right|M(p)\right\rangle $,
and care should be drawn when expressing this quantity in terms of
the decay constant. Since the decay width of the meson $\Gamma\sim|g_{L}^{ij}|^{2}+|g_{R}^{ij}|^{2}$,
we can parameterize the FV matrix element as follows: 
\begin{equation}
\left\langle 0\left|J_{Z,\text{FV}}^{\mu}\right|M(p)\right\rangle \equiv\frac{\sqrt{|g_{L}^{ij}|^{2}+|g_{R}^{ij}|^{2}}}{\sqrt{|g_{L}^{q}|^{2}+|g_{R}^{q}|^{2}}}\left\langle 0\left|J_{Z}^{\mu}\right|M(p)\right\rangle ,\label{eq:decay_const}
\end{equation}
where $g_{L,R}^{q}$ are the SM couplings. This parameterization is
quite plausible, as if FV indeed exists, it must contribute to the
(experimentally measured) decay constant. In addition, the SM limit
is restored simply by setting $i=j$, as it should. With this parameterization,
the decay width in Figure \ref{fig2} can easily be evaluated in terms
of the decay constant of the decaying meson.

We start with vector mesons, which are mesons that have the quantum
number $J^{PC}=1^{--}$, These mesons are created from QCD vacuum
via the following matrix element 
\begin{equation}
\left\langle 0\left|J_{Z,\text{FV}}^{\mu}\right|M(p)\right\rangle =f_{V}m_{V}\epsilon^{\mu}(p),\label{eq:Vector_QCD}
\end{equation}
where $m_{V}$ is the mass of the vector meson and $f_{V}$ is its
decay constant. The decay width for vector mesons is given by 
\begin{equation}
\Gamma(V \to \ell\ell)=\frac{f_{V}^{2}m_{V}^{3}}{24\pi M_{Z}^{4}}\Big(\frac{|g_{L}^{ij}|^{2}+|g_{R}^{ij}|^{2}}{|g_{L}^{q}|^{2}+|g_{R}^{q}|^{2}}\Big)\left(g_{L}^{\ell2}+g_{R}^{\ell2}-\frac{m_{\ell}^{2}}{m_{V}^{2}}\left(g_{L}^{\ell2}-6g_{L}^{\ell}g_{R}^{\ell}+g_{R}^{\ell2}\right)\right)\left(1-\frac{m_{V}^{2}}{M_{Z}^{2}}\right)^{-2}\sqrt{1-\frac{4m_{\ell}^{2}}{m_{V}^{2}}},
\end{equation}
however, the only neutral vector meson that is made of $q_{i}\bar{q}_{j}$
with $i\neq j$ is $K_{0}^{*}$, and it is not known that this meson
decays leptonically. While no dedicated negative results are known
to exist for $K_{0}^{*} \to \ell^{+}\ell^{-}$, as this decay is expected
to be unobservably rare in the SM, future high-precision experiments
might probe this channel.

We turn our attention to pseudoscalar mesons, which are mesons with
the quantum numbers $J^{PC}=0^{-+}$. These mesons are created from
the QCD vacuum through the matrix element 
\begin{equation}
\left\langle 0\left|J_{Z,\text{FV}}^{\mu}\right|P(p)\right\rangle =-if_{p}p^{\mu},\label{eq:ps_QCD}
\end{equation}
which implies that the decay width is given by 
\begin{equation}
\Gamma(P \to \ell\ell)=\frac{g^{2}f_{P}^{2}}{32\pi\cos^{2}{\theta_{W}}}\Big(\frac{|g_{L}^{ij}|^{2}+|g_{R}^{ij}|^{2}}{|g_{L}^{q}|^{2}+|g_{R}^{q}|^{2}}\Big)\frac{m_{P}m_{l}^{2}}{M_{Z}^{4}}\sqrt{1-\frac{4m_{l}^{2}}{M_{Z}^{2}}},
\end{equation}
where $m_{P}$ is the mass of the pseudoscalar meson, and $f_{P}$
is its decay constant. The neutral mesons made of quarks of different
species which decay leptonically include $D^{0}$, $B^{0}$, $B_{s}$,
$K_{s}^{0}$ and $K_{L}^{0}$ mesons.

All bounds listed here are taken from~\cite{ParticleDataGroup:2024cfk},
except for $B^{0} \to \mu^{+}\mu^{-}$, which is from~\cite{CMS:2022mgd}.
For $D^{0}$ decays, the latest bounds are given by $\text{Br}(D^{0} \to \mu^{+}\mu^{-})<3.1\times10^{-9}$
and $\text{Br}(D^{0} \to e^{+}e^{-})<7.9\times10^{-8}$, which translate
to $\sqrt{|g_{L}^{cu}|^{2}+|g_{R}^{cu}|^{2}}<7.44\times10^{-5}$ and
$\sqrt{|g_{L}^{cu}|^{2}+|g_{R}^{cu}|^{2}}<7.76\times10^{-2}$, respectively.
For $B^{0}$ decays, the bounds read $\text{Br}(B^{0} \to \tau^{+}\tau^{-})<2.1\times10^{-3}$,
$\text{Br}(B^{0} \to \mu^{+}\mu^{-})<1.5\times10^{-10}$ and $\text{Br}(B^{0} \to e^{+}e^{-})<2.5\times10^{-9}$.
These bounds yield $\sqrt{|g_{L}^{bd}|^{2}+|g_{R}^{bd}|^{2}}<1.5\times10^{-3}$,
$\sqrt{|g_{L}^{bd}|^{2}+|g_{R}^{bd}|^{2}}<5.79\times10^{-6}$ and
$\sqrt{|g_{L}^{bd}|^{2}+|g_{R}^{bd}|^{2}}<5.4\times10^{-3}$, respectively.
Bounds on FV in the decays of $B_{s}$ mesons are given by $\text{Br}(B_{s} \to \tau^{+}\tau^{-})<6.8\times10^{-3}$,
$\text{Br}(B_{s} \to \mu^{+}\mu^{-})=(3.01\pm0.35)\times10^{-9}$ and
$\text{Br}(B_{s} \to e^{+}e^{-})<9.4\times10^{-9}$. These translate
to $\sqrt{|g_{L}^{bs}|^{2}+|g_{R}^{bs}|^{2}}<2.2\times10^{-3}$, $\sqrt{|g_{L}^{bs}|^{2}+|g_{R}^{bs}|^{2}}<1.03\times10^{-5}$
and $\sqrt{|g_{L}^{bs}|^{2}+|g_{R}^{bs}|^{2}}<8.6\times10^{-3}$,
respectively. Bounds involving $K_{s}^{0}$ decays are given by $\text{Br}(K_{s}^{0} \to \mu^{+}\mu^{-})<2.1\times10^{-10}$
and $\text{Br}(K_{s}^{0} \to e^{+}e^{-})<9\times10^{-9}$, which yield
$\sqrt{|g_{L}^{sd}|^{2}+|g_{R}^{sd}|^{2}}<4.04\times10^{-6}$ and
$\sqrt{|g_{L}^{sd}|^{2}+|g_{R}^{sd}|^{2}}<5.22\times10^{-3}$, respectively.
Finally, bounds involving $K_{L}^{0}$ decays are given by $\text{Br}(K_{s}^{0} \to \mu^{+}\mu^{-})=(6.84\pm0.11)\times10^{-9}$
and $\text{Br}(K_{s}^{0} \to e^{+}e^{-})=(9\substack{+6\\
-4
}
)\times10^{-12}$, which give the bounds $\sqrt{|g_{L}^{sd}|^{2}+|g_{R}^{sd}|^{2}}<4.96\times10^{-7}$
and $\sqrt{|g_{L}^{sd}|^{2}+|g_{R}^{sd}|^{2}}<7.23\times10^{-6}$,
respectively.

\begin{figure}[ht!]
\includegraphics[width=0.5\textwidth]{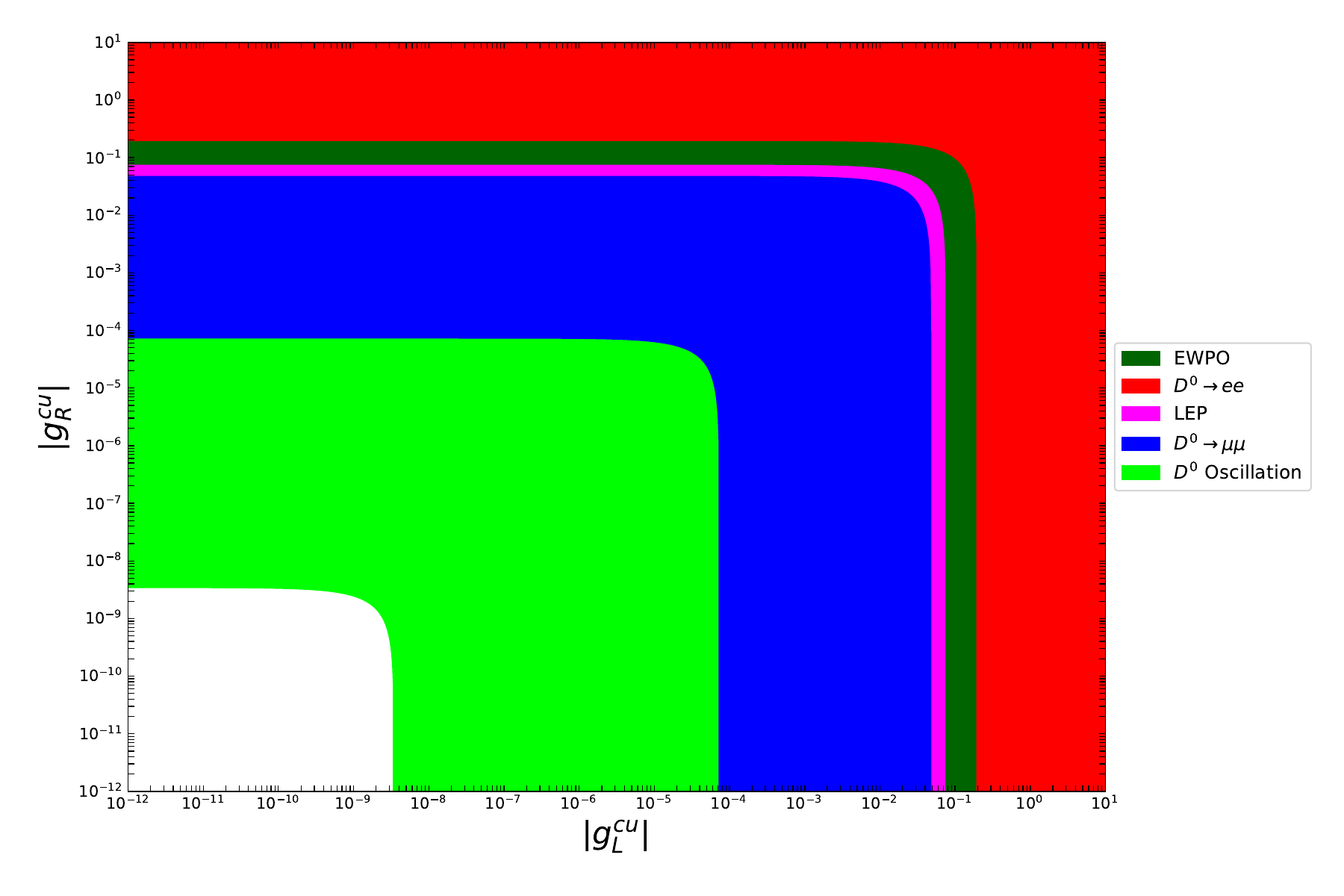}~\includegraphics[width=0.5\textwidth]{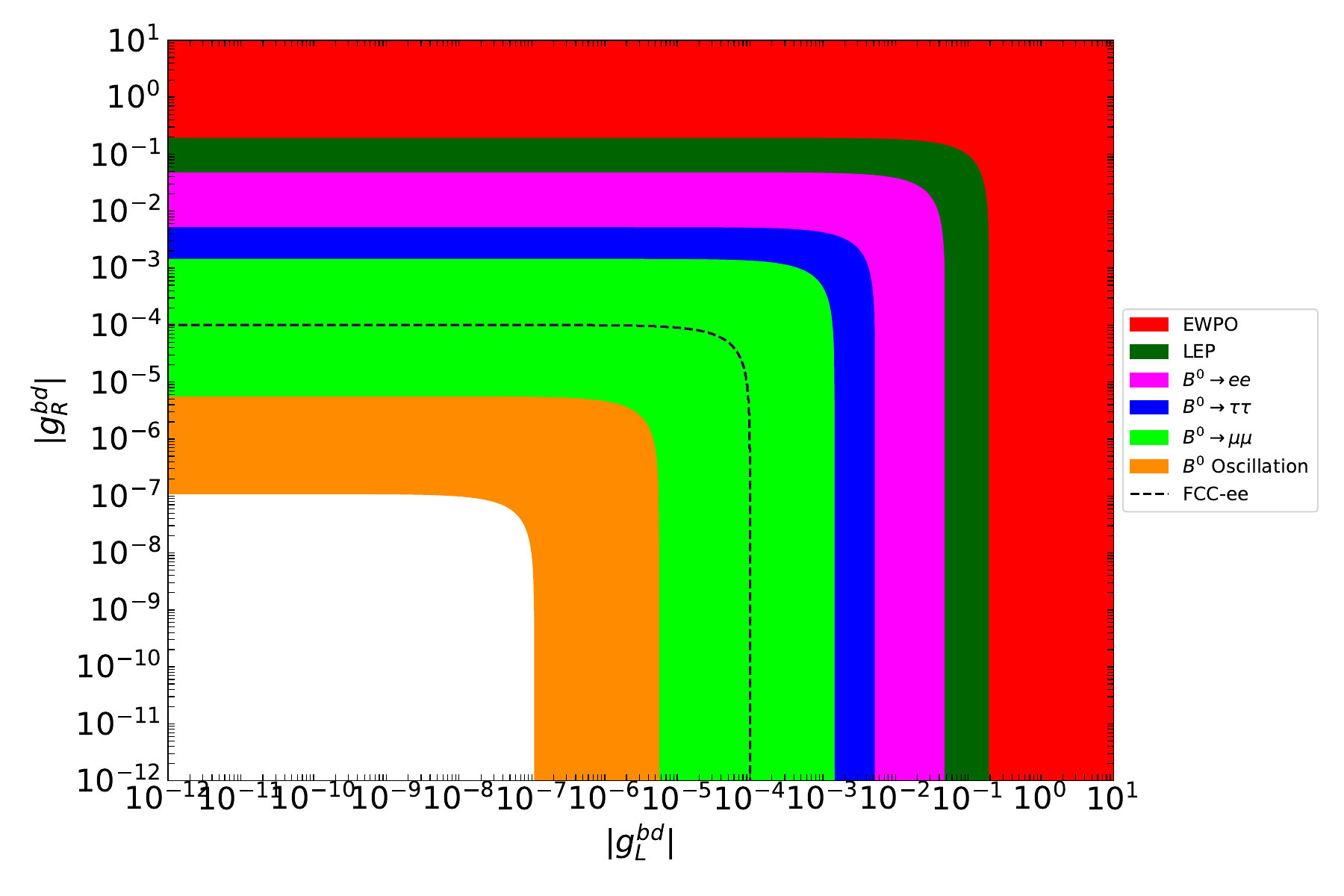}\\
 \includegraphics[width=0.5\textwidth]{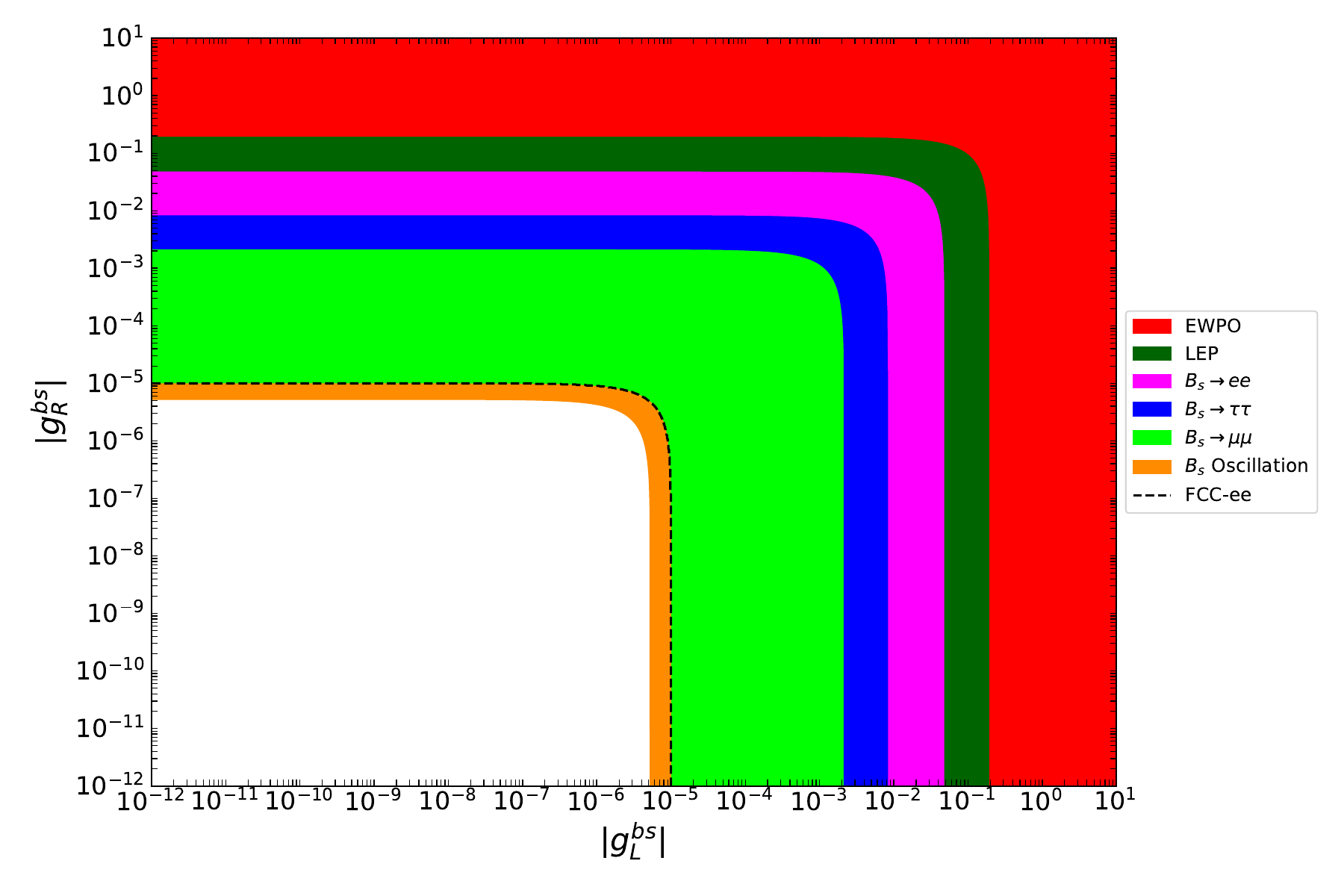}~\includegraphics[width=0.5\textwidth]{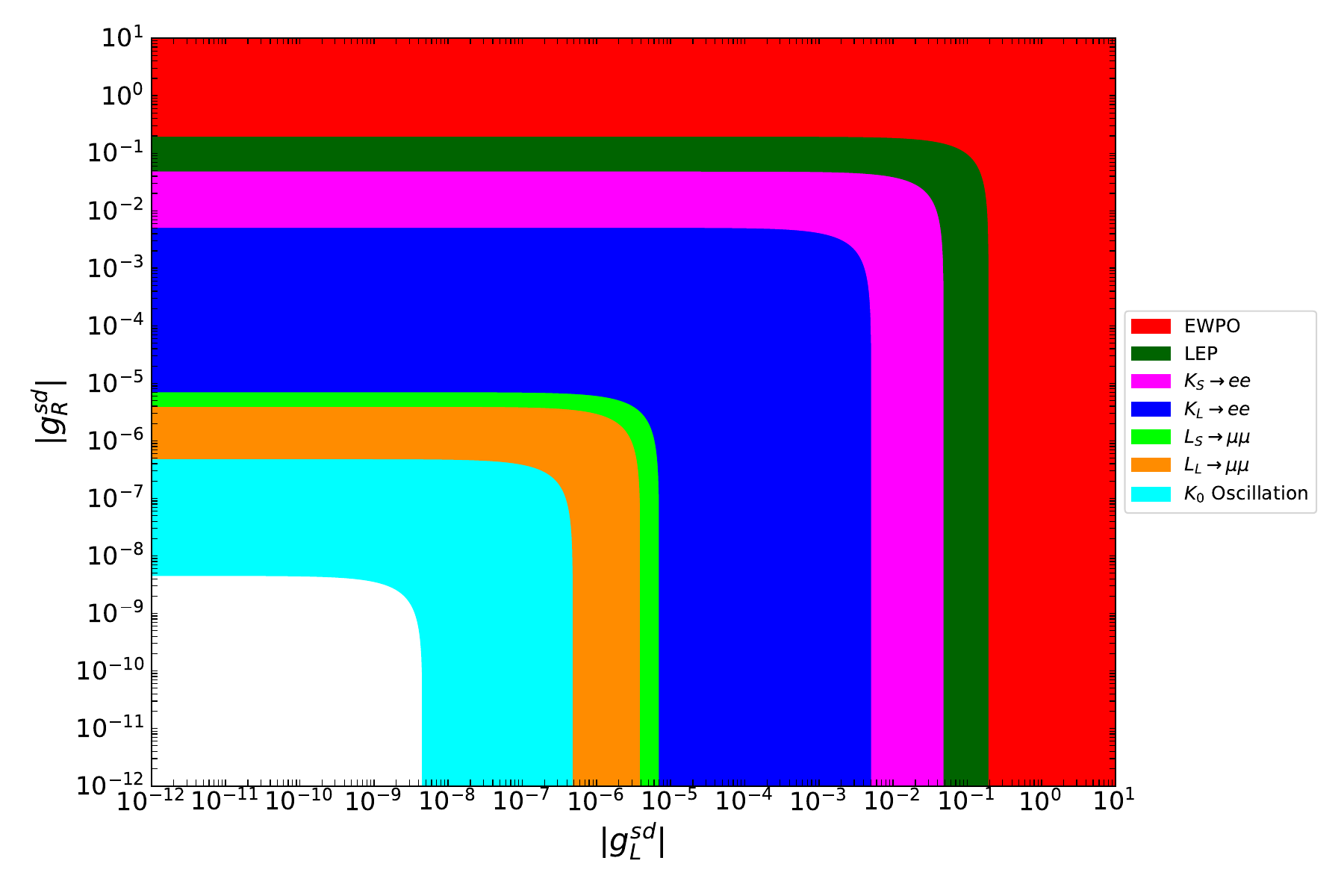}\\
 \includegraphics[width=0.5\textwidth]{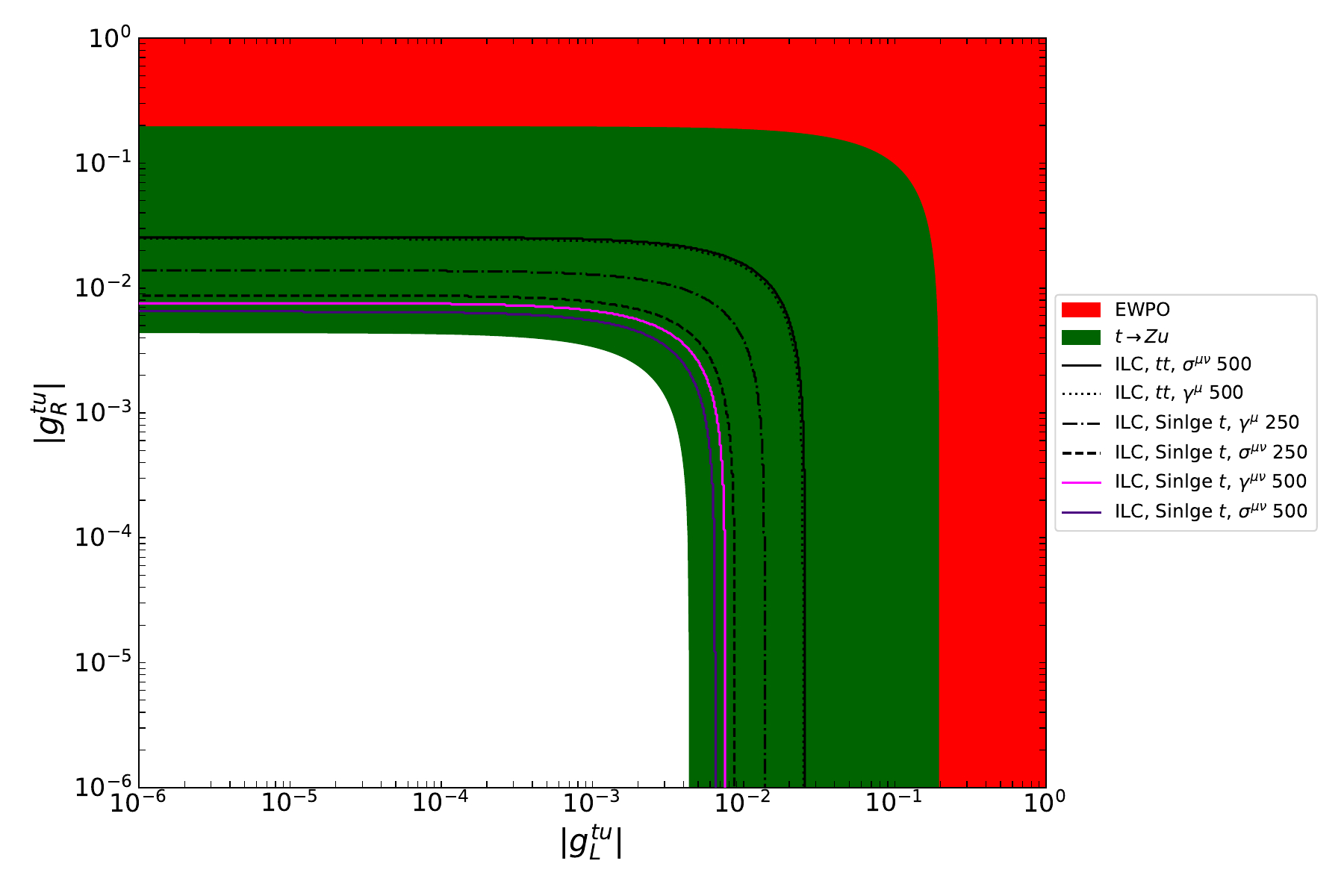}~\includegraphics[width=0.5\textwidth]{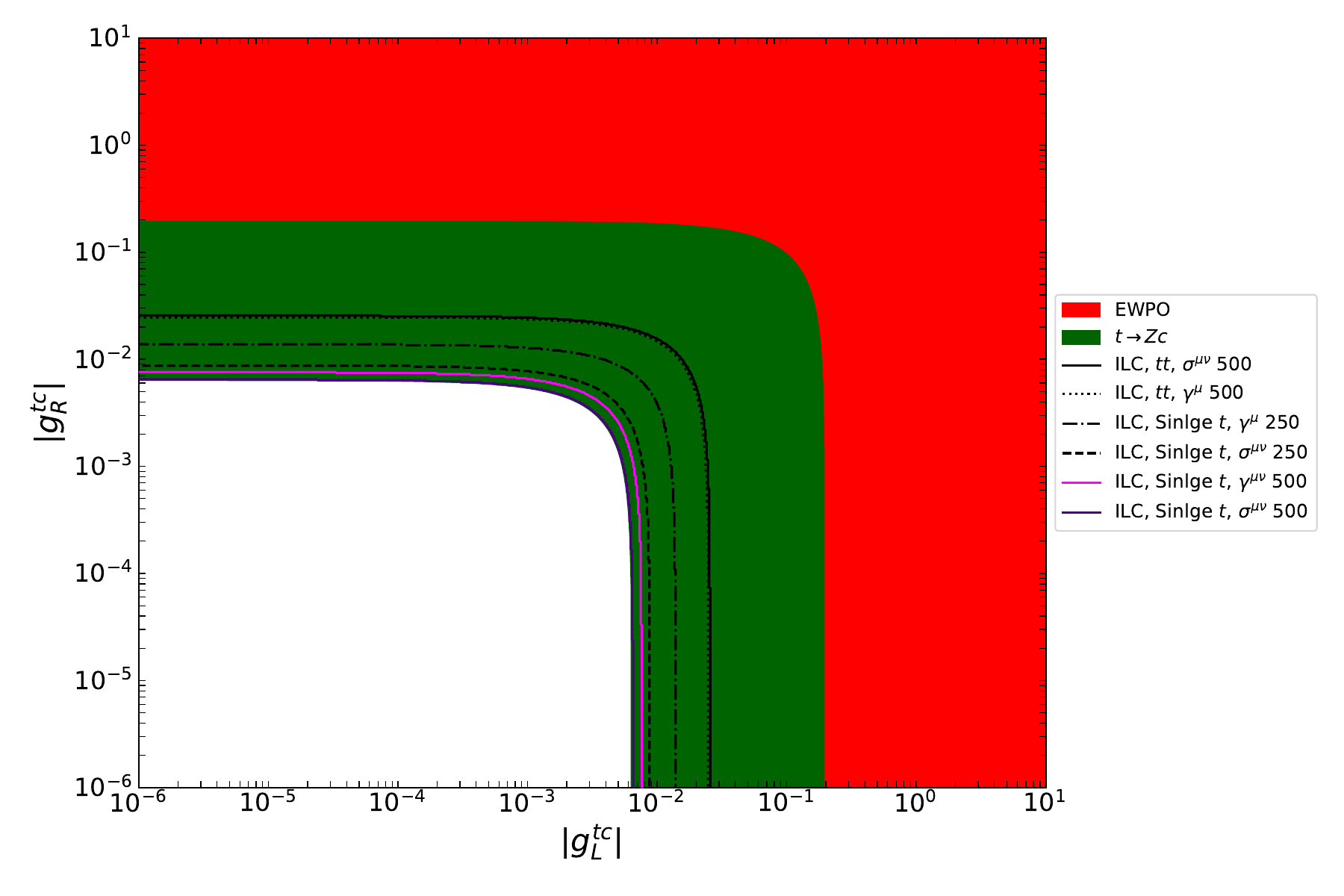}
\caption{The current bounds and future projections of the FV $Z$ couplings
to quarks ($|g_{L,R}^{ij}|$) given in Eq.~(\ref{eq:FV_Z_cpoupling}). }
\label{fig3} 
\end{figure}

Before concluding this section, we briefly note that it is possible
to construct "mixed" bounds involving FV couplings in both the quark
and lepton sectors by considering decays such as $D^{0} \to \mu e$.
However, due to the extremely stringent existing constraints on FV
$Z$ couplings to leptons (see~\cite{Abu-Ajamieh:2025vxw} for details),
the corresponding limits on the quark sector in such channels are
relatively weak. More specifically, such mixed decays have 
\[
\Gamma\sim\left(|g_{L}^{q_{i}q_{j}}|^{2}+|g_{R}^{q_{i}q_{j}}|^{2}\right)\left(|g_{L}^{\ell_{i}\ell_{j}}|^{2}+|g_{R}^{\ell_{i}\ell_{j}}|^{2}\right),
\]
and we have checked numerically that, if we use the upper limits on
the relevant FV leptonic couplings found in Ref.~\cite{Abu-Ajamieh:2025vxw}
to set conservative bounds on $|g_{L}^{q_{i}q_{j}}|^{2}+|g_{R}^{q_{i}q_{j}}|^{2}$,
the resulting bounds are indeed very weak. A complete summary of all
bounds and projections is provided in Figure~\ref{fig3}.

\section{Conclusions and Outlook~\label{sec:conclusion}}

In this paper, we investigated the current bounds and future projections
on flavor-violating (FV) couplings of the $Z$ boson to quarks. In
particular, we analyzed existing experimental constraints from electroweak
precision observables (EWPO), meson oscillations, rare top quark decays
($t \to Zq$), and the decays of neutral mesons composed of quarks
of different flavors. Our findings indicate that the EWPO and direct
collider searches are the least sensitive to these couplings, with
bounds typically at the level of $\sim10^{-1}$--$10^{-2}$. Constraints
from top decays, which probe $Z$ couplings to $tu$ and $tc$, are
more stringent, reaching $\sim10^{-3}$. Notably, these current limits
are stronger than the most optimistic projections from future searches
at the ILC.

We found that bounds from neutral meson decays span a range from $\sim10^{-2}$
to $10^{-7}$, while meson oscillation measurements provide the strongest
constraints, reaching sensitivities as low as $\sim10^{-6}$--$10^{-9}$.
These results clearly demonstrate that low-energy experiments are
more effective in probing FV couplings than high-energy collider searches.
This dramatic disparity underscores the unique sensitivity of flavor
observables to FV effects. Consequently, we advocate for greater experimental
focus on low-energy flavor observables. Furthermore, flavor violation
in the quark sector has historically received less attention than
that in the lepton sector, both for Higgs and $Z$ couplings. The
quark sector’s hierarchical Yukawa couplings and strong QCD backgrounds
make FV searches both challenging and theoretically informative. We
urge enhanced efforts in both theory and experiment to probe quark-sector
FV.

\section*{Acknowledgment}

The work of AA is supported by Sharjah University via the HEP research
group operational grant. SK would like to acknowledge for financial
support through the ANRF National Postdoctoral Fellowship (NPDF) with
project grant no PDF/2023/000410. The work of NO is supported in part
by the United States Department of Energy Grant, Nos. DE-SC0012447
and DC-SC0023713.


\begin{thebibliography}{10}

\bibitem{ATLAS:2012yve} G.~Aad \textit{et al.} [ATLAS], ``Observation
of a new particle in the search for the Standard Model Higgs boson
with the ATLAS detector at the LHC,'' Phys. Lett. B \textbf{716},
1-29 (2012) \arXivold{1207.7214}{hep-ex}. 

\bibitem{CMS:2012qbp} S.~Chatrchyan \textit{et al.} [CMS], ``Observation
of a New Boson at a Mass of 125 GeV with the CMS Experiment at the
LHC,'' Phys. Lett. B \textbf{716}, 30-61 (2012) \arXivold{1207.7235}{hep-ex}.

\bibitem{Nefediev:2025vmo}
A.~Nefediev,
Int. J. Mod. Phys. A \textbf{40}, no.34, 2530015 (2025) 
[arXiv:2507.19256 [hep-ph]].

\bibitem{Buchmuller:1985jz} W.~Buchmuller and D.~Wyler, ``Effective
Lagrangian Analysis of New Interactions and Flavor Conservation,''
Nucl. Phys. B \textbf{268}, 621-653 (1986) 

\bibitem{Grzadkowski:2010es} B.~Grzadkowski, M.~Iskrzynski, M.~Misiak
and J.~Rosiek, ``Dimension-Six Terms in the Standard Model Lagrangian,''
JHEP \textbf{10}, 085 (2010) \arXivold{1008.4884}{hep-ph}. 

\bibitem{Brivio:2017vri} I.~Brivio and M.~Trott, ``The Standard
Model as an Effective Field Theory,'' Phys. Rept. \textbf{793}, 1-98
(2019) \arXivold{1706.08945}{hep-ph}. 

\bibitem{Chang:2019vez} S.~Chang and M.~A.~Luty, ``The Higgs
Trilinear Coupling and the Scale of New Physics,'' JHEP \textbf{03}
(2020), 140 \arXivold{1902.05556}{hep-ph}. 

\bibitem{Abu-Ajamieh:2020yqi} F.~Abu-Ajamieh, S.~Chang, M.~Chen
and M.~A.~Luty, ``Higgs coupling measurements and the scale of
new physics,'' JHEP \textbf{07}, 056 (2021) \arXivold{2009.11293}{hep-ph}.

\bibitem{Abu-Ajamieh:2021egq} F.~Abu-Ajamieh, ``The scale of new
physics from the Higgs couplings to \ensuremath{\gamma}\ensuremath{\gamma}
and \ensuremath{\gamma}Z,'' JHEP \textbf{06}, 091 (2022) \arXivold{2112.13529}{hep-ph}.

\bibitem{Abu-Ajamieh:2022ppp} F.~Abu-Ajamieh, ``The scale of new
physics from the Higgs couplings to gg,'' Phys. Lett. B \textbf{833},
137389 (2022) \arXivold{2203.07410}{hep-ph}. 

\bibitem{Grober:2025eyl}
R.~Gr{\"o}ber,
[arXiv:2512.01499 [hep-ph]].

\bibitem{Glashow:1970gm} S.~L.~Glashow, J.~Iliopoulos and L.~Maiani,
``Weak Interactions with Lepton-Hadron Symmetry,'' Phys. Rev. D
\textbf{2} (1970), 1285-1292 

\bibitem{Foot:1992ui} R.~Foot, H.~Lew and R.~R.~Volkas, ``Electric
charge quantization,'' J. Phys. G \textbf{19}, 361-372 (1993) [erratum:
J. Phys. G \textbf{19}, 1067 (1993)] \arXivold{hep-ph/9209259}{hep-ph}.

\bibitem{Giunti:2014ixa} C.~Giunti and A.~Studenikin, ``Neutrino
electromagnetic interactions: a window to new physics,'' Rev. Mod.
Phys. \textbf{87}, 531 (2015) \arXivold{1403.6344}{hep-ph}. 

\bibitem{Raffelt:1999gv} G.~G.~Raffelt, ``Limits on neutrino electromagnetic
properties: An update,'' Phys. Rept. \textbf{320}, 319-327 (1999)

\bibitem{Abu-Ajamieh:2023txh} F.~Abu-Ajamieh, N.~Okada and S.~K.~Vempati,
``Corrected calculation for the non-local solution to the g \ensuremath{-}
2 anomaly and novel results in non-local QED,'' JHEP \textbf{01},
015 (2024) \arXivold{2309.08417}{hep-ph}. 

\bibitem{Bjorken:1977vt} J.~D.~Bjorken and S.~Weinberg, ``A Mechanism
for Nonconservation of Muon Number,'' Phys. Rev. Lett. \textbf{38},
622 (1977) 

\bibitem{McWilliams:1980kj} B.~McWilliams and L.~F.~Li, ``Virtual
Effects of Higgs Particles,'' Nucl. Phys. B \textbf{179}, 62-84 (1981)

\bibitem{Shanker:1981mj} O.~U.~Shanker, ``Flavor Violation, Scalar
Particles and Leptoquarks,'' Nucl. Phys. B \textbf{206}, 253-272
(1982) 

\bibitem{Barr:1990vd} S.~M.~Barr and A.~Zee, ``Electric Dipole
Moment of the Electron and of the Neutron,'' Phys. Rev. Lett. \textbf{65},
21-24 (1990) [erratum: Phys. Rev. Lett. \textbf{65}, 2920 (1990)]

\bibitem{Babu:1999me} K.~S.~Babu and S.~Nandi, ``Natural fermion
mass hierarchy and new signals for the Higgs boson,'' Phys. Rev.
D \textbf{62}, 033002 (2000) \arXivold{hep-ph/9907213}{hep-ph}.

\bibitem{Diaz-Cruz:1999sns} J.~L.~Diaz-Cruz and J.~J.~Toscano,
``Lepton flavor violating decays of Higgs bosons beyond the standard
model,'' Phys. Rev. D \textbf{62}, 116005 (2000) \arXivold{hep-ph/9910233}{hep-ph}.

\bibitem{Han:2000jz} T.~Han and D.~Marfatia, ``h ---\ensuremath{>}
mu tau at hadron colliders,'' Phys. Rev. Lett. \textbf{86}, 1442-1445
(2001) \arXivold{hep-ph/0008141}{hep-ph}. 

\bibitem{Blanke:2008zb} M.~Blanke, A.~J.~Buras, B.~Duling, S.~Gori
and A.~Weiler, ``$\Delta$ F=2 Observables and Fine-Tuning in a
Warped Extra Dimension with Custodial Protection,'' JHEP \textbf{03},
001 (2009) \arXivold{0809.1073}{hep-ph}. 

\bibitem{Casagrande:2008hr} S.~Casagrande, F.~Goertz, U.~Haisch,
M.~Neubert and T.~Pfoh, ``Flavor Physics in the Randall-Sundrum
Model: I. Theoretical Setup and Electroweak Precision Tests,'' JHEP
\textbf{10}, 094 (2008) \arXivold{0807.4937}{hep-ph}. 

\bibitem{Giudice:2008uua} G.~F.~Giudice and O.~Lebedev, ``Higgs-dependent
Yukawa couplings,'' Phys. Lett. B \textbf{665}, 79-85 (2008) \arXivold{0804.1753}{hep-ph}.

\bibitem{Aguilar-Saavedra:2009ygx} J.~A.~Aguilar-Saavedra, ``A
Minimal set of top-Higgs anomalous couplings,'' Nucl. Phys. B \textbf{821},
215-227 (2009) \arXivold{0904.2387}{hep-ph}. 

\bibitem{Albrecht:2009xr} M.~E.~Albrecht, M.~Blanke, A.~J.~Buras,
B.~Duling and K.~Gemmler, ``Electroweak and Flavour Structure of
a Warped Extra Dimension with Custodial Protection,'' JHEP \textbf{09},
064 (2009) \arXivold{0903.2415}{hep-ph}. 

\bibitem{Buras:2009ka} A.~J.~Buras, B.~Duling and S.~Gori, ``The
Impact of Kaluza-Klein Fermions on Standard Model Fermion Couplings
in a RS Model with Custodial Protection,'' JHEP \textbf{09}, 076
(2009) \arXivold{0905.2318}{hep-ph}. 

\bibitem{Agashe:2009di} K.~Agashe and R.~Contino, ``Composite
Higgs-Mediated FCNC,'' Phys. Rev. D \textbf{80}, 075016 (2009) \arXivold{0906.1542}{hep-ph}.

\bibitem{Goudelis:2011un} A.~Goudelis, O.~Lebedev and J.~h.~Park,
``Higgs-induced lepton flavor violation,'' Phys. Lett. B \textbf{707},
369-374 (2012) \arXivold{1111.1715}{hep-ph}. 

\bibitem{Arhrib:2012mg} A.~Arhrib, Y.~Cheng and O.~C.~W.~Kong,
``Higgs to mu+tau Decay in Supersymmetry without R-parity,'' EPL
\textbf{101}, no.3, 31003 (2013) \arXivold{1208.4669}{hep-ph}.

\bibitem{McKeen:2012av} D.~McKeen, M.~Pospelov and A.~Ritz, ``Modified
Higgs branching ratios versus CP and lepton flavor violation,'' Phys.
Rev. D \textbf{86}, 113004 (2012) \arXivold{1208.4597}{hep-ph}.

\bibitem{Azatov:2009na} A.~Azatov, M.~Toharia and L.~Zhu, ``Higgs
Mediated FCNC's in Warped Extra Dimensions,'' Phys. Rev. D \textbf{80},
035016 (2009) \arXivold{0906.1990}{hep-ph}. 

\bibitem{Blankenburg:2012ex} G.~Blankenburg, J.~Ellis and G.~Isidori,
``Flavour-Changing Decays of a 125 GeV Higgs-like Particle,'' Phys.
Lett. B \textbf{712}, 386-390 (2012) \arXivold{1202.5704}{hep-ph}.

\bibitem{Kanemura:2005hr} S.~Kanemura, T.~Ota and K.~Tsumura,
``Lepton flavor violation in Higgs boson decays under the rare tau
decay results,'' Phys. Rev. D \textbf{73}, 016006 (2006) \arXivold{hep-ph/0505191}{hep-ph}.

\bibitem{Davidson:2010xv} S.~Davidson and G.~J.~Grenier, ``Lepton
flavour violating Higgs and tau to mu gamma,'' Phys. Rev. D \textbf{81},
095016 (2010) \arXivold{1001.0434}{hep-ph}. 

\bibitem{Harnik:2012pb} R.~Harnik, J.~Kopp and J.~Zupan, ``Flavor
Violating Higgs Decays,'' JHEP \textbf{03}, 026 (2013) \arXivold{1209.1397}{hep-ph}.

\bibitem{Fernandez:2009vr} A.~Fernandez, C.~Pagliarone, F.~Ramirez-Zavaleta
and J.~J.~Toscano, ``Higgs mediated Double Flavor Violating top
decays in Effective Theories,'' J. Phys. G \textbf{37}, 085007 (2010)
\arXivold{0911.4995}{hep-ph}. 

\bibitem{Aranda:2009cd} J.~I.~Aranda, A.~Cordero-Cid, F.~Ramirez-Zavaleta,
J.~J.~Toscano and E.~S.~Tututi, ``Higgs mediated flavor violating
top quark decays $t \to u_{i}H$, $u_{i}\gamma$, $u_{i}\gamma\gamma$,
and the process $\gamma\gamma \to tc$ in effective theories,'' Phys.
Rev. D \textbf{81}, 077701 (2010) \arXivold{0911.2304}{hep-ph}.

\bibitem{Aranda:2010qc} J.~I.~Aranda, J.~Montano, F.~Ramirez-Zavaleta,
J.~J.~Toscano and E.~S.~Tututi, ``Bounding the $B_{s}\rightarrow\gamma\gamma$
decay from Higgs mediated FCNC transitions,'' Phys. Rev. D \textbf{82},
054002 (2010) \arXivold{1005.5452}{hep-ph}. 

\bibitem{Abu-Ajamieh:2022nmt} F.~Abu-Ajamieh and S.~K.~Vempati,
``Can the Higgs still account for the g\ensuremath{-}2 anomaly?,''
Int. J. Mod. Phys. A \textbf{38}, no.20, 2350091 (2023) \arXivold{2209.10898}{hep-ph}.

\bibitem{Abu-Ajamieh:2023qvh} F.~Abu-Ajamieh, M.~Frasca and S.~K.~Vempati,
``Flavor Violating Di- Higgs Coupling,'' \arXivold{2305.17362}{hep-ph}.

\bibitem{Abu-Ajamieh:2025jsz} F.~Abu-Ajamieh, S.~Kumbhakar, R.~Sarkar
and S.~Vempati, ``Improved Bounds and Global Fit of Flavor-Violating
Charged Lepton Yukawa Couplings post LHC,'' \arXivold{2505.12208}{hep-ph}.

\bibitem{Brignole:2004ah} A.~Brignole and A.~Rossi, ``Anatomy
and phenomenology of mu-tau lepton flavor violation in the MSSM,''
Nucl. Phys. B \textbf{701}, 3-53 (2004) \arXivold{hep-ph/0404211}{hep-ph}.

\bibitem{Davidson:2012wn} S.~Davidson, S.~Lacroix and P.~Verdier,
``LHC sensitivity to lepton flavour violating Z boson decays,''
JHEP \textbf{09}, 092 (2012) \arXivold{1207.4894}{hep-ph}. 

\bibitem{Goto:2015iha} T.~Goto, R.~Kitano and S.~Mori, ``Lepton
flavor violating $Z$-boson couplings from nonstandard Higgs interactions,''
Phys. Rev. D \textbf{92}, 075021 (2015) \arXivold{1507.03234}{hep-ph}.

\bibitem{Kamenik:2023hvi} J.~F.~Kamenik, A.~Korajac, M.~Szewc,
M.~Tammaro and J.~Zupan, ``Flavor-violating Higgs and Z boson decays
at a future circular lepton collider,'' Phys. Rev. D \textbf{109},
no.1, L011301 (2024) \arXivold{2306.17520}{hep-ph}. 

\bibitem{Jueid:2023fgo} A.~Jueid, J.~Kim, S.~Lee, J.~Song and
D.~Wang, ``Exploring lepton flavor violation phenomena of the Z
and Higgs bosons at electron-proton colliders,'' Phys. Rev. D \textbf{108},
no.5, 055024 (2023) \arXivold{2305.05386}{hep-ph}. 

\bibitem{Calibbi:2021pyh} L.~Calibbi, X.~Marcano and J.~Roy, ``Z
lepton flavour violation as a probe for new physics at future $e^{+}e^{-}$
colliders,'' Eur. Phys. J. C \textbf{81}, no.12, 1054 (2021) \arXivold{2107.10273}{hep-ph}.

\bibitem{Chivukula:2002ry} R.~S.~Chivukula and E.~H.~Simmons,
``Electroweak limits on nonuniversal Z-prime bosons,'' Phys. Rev.
D \textbf{66}, 015006 (2002) \arXivold{hep-ph/0205064}{hep-ph}.

\bibitem{Erler:2009jh} J.~Erler, P.~Langacker, S.~Munir and E.~Rojas,
``Improved Constraints on Z-prime Bosons from Electroweak Precision
Data,'' JHEP \textbf{08}, 017 (2009) \arXivold{0906.2435}{hep-ph}.

\bibitem{Aranda:2010cy} J.~I.~Aranda, F.~Ramirez-Zavaleta, J.~J.~Toscano
and E.~S.~Tututi, ``Bounding the $Z^{\prime}tc$ coupling from
$D^{0}-\overline{D^{0}}$ mixing and single top production at the
ILC,'' J. Phys. G \textbf{38}, 045006 (2011) \arXivold{1007.3326}{hep-ph}.

\bibitem{Murakami:2001cs} B.~Murakami, ``The Impact of lepton flavor
violating Z-prime bosons on muon g-2 and other muon observables,''
Phys. Rev. D \textbf{65}, 055003 (2002) \arXivold{hep-ph/0110095}{hep-ph}.

\bibitem{Altmannshofer:2016brv} W.~Altmannshofer, C.~Y.~Chen,
P.~S.~Bhupal Dev and A.~Soni, ``Lepton flavor violating Z' explanation
of the muon anomalous magnetic moment,'' Phys. Lett. B \textbf{762},
389-398 (2016) \arXivold{1607.06832}{hep-ph}. 

\bibitem{Abu-Ajamieh:2025vxw} F.~Abu-Ajamieh, A.~Ahriche and N.~Okada,
``Novel and Updated Bounds on Flavor-violating Z Interactions in
the Lepton Sector,'' \arXivold{2503.07236}{hep-ph}. 

\bibitem{OPAL:2000ufp} G.~Abbiendi \textit{et al.} [OPAL], ``Precise
determination of the Z resonance parameters at LEP: 'Zedometry',''
Eur. Phys. J. C \textbf{19}, 587-651 (2001) \arXivold{hep-ex/0012018}{hep-ex}.

\bibitem{UTfit:2007eik} M.~Bona \textit{et al.} [UTfit], ``Model-independent
constraints on $\Delta F=2$ operators and the scale of new physics,''
JHEP \textbf{03}, 049 (2008) \arXivold{0707.0636}{hep-ph}. 

\bibitem{ATLAS:2023qzr} G.~Aad \textit{et al.} [ATLAS], ``Search
for flavor-changing neutral-current couplings between the top quark
and the Z boson with proton-proton collisions at s=13\,\,TeV with
the ATLAS detector,'' Phys. Rev. D \textbf{108}, no.3, 032019 (2023)
\arXivold{2301.11605}{hep-ex}. 

\bibitem{D0:2012hgn} V.~M.~Abazov \textit{et al.} [D0], ``An
Improved determination of the width of the top quark,'' Phys. Rev.
D \textbf{85}, 091104 (2012) \arXivold{1201.4156}{hep-ex}. 

\bibitem{Degrassi:2012ry} G.~Degrassi, S.~Di Vita, J.~Elias-Miro,
J.~R.~Espinosa, G.~F.~Giudice, G.~Isidori and A.~Strumia, ``Higgs
mass and vacuum stability in the Standard Model at NNLO,'' JHEP \textbf{08},
098 (2012) \arXivold{1205.6497}{hep-ph}. 

\bibitem{TopQuarkWorkingGroup:2013hxj} K.~Agashe \textit{et al.}
[Top Quark Working Group], ``Working Group Report: Top Quark,''
\arXivold{1311.2028}{hep-ph}. 

\bibitem{ParticleDataGroup:2024cfk} S.~Navas \textit{et al.} [Particle
Data Group], ``Review of particle physics,'' Phys. Rev. D \textbf{110}
(2024) no.3, 030001 

\bibitem{CMS:2022mgd} A.~Tumasyan \textit{et al.} [CMS], ``Measurement
of the B$_{s}^{0} \to \mu^{+}\mu^{-}$ decay properties and search for
the B$^{0} \to \mu^{+}\mu^{-}$ decay in proton-proton collisions at
$\sqrt{s}$ = 13 TeV,'' Phys. Lett. B \textbf{842}, 137955 (2023)
\arXivold{2212.10311}{hep-ex}. 

\end{thebibliography}
\end{document}